\documentclass[pra,preprint,superscriptaddress,floatfix,showpacs]{revtex4-1}

\usepackage{amsmath,amsfonts,amssymb}
\usepackage{graphicx}

\def\beq{\begin{equation}}
\def\eeq{\end{equation}}
\def\beqn{\begin{eqnarray}}
\def\eeqn{\end{eqnarray}}

\def\bl {\mbox{\boldmath $[$}}
\def\br {\mbox{\boldmath $]$}}
\def\r {{\bf r}}
\def\p {{\bf p}}
\def\Q {{\bf Q}}

\begin{document}

\title{The uncertainty product of an out-of-equilibrium\\ many-particle system}
\author{Shachar Klaiman}
\email{shachar.klaiman@pci.uni-heidelberg.de}
\affiliation{Theoretische Chemie, Physikalisch--Chemisches Institut, Universit\"at Heidelberg, 
Im Neuenheimer Feld 229, D-69120 Heidelberg, Germany}
\author{Alexej I. Streltsov}
\email{alexej.streltsov@pci.uni-heidelberg.de}
\affiliation{Theoretische Chemie, Physikalisch--Chemisches Institut, Universit\"at Heidelberg, 
Im Neuenheimer Feld 229, D-69120 Heidelberg, Germany}
\author{Ofir E. Alon}
\email{ofir@research.haifa.ac.il}
\affiliation{Department of Physics, University of Haifa at Oranim, Tivon 36006, Israel}
\date{\today}

\begin{abstract}

In the present work we show, analytically and numerically, that the variance of many-particle operators and
their uncertainty product for an out-of-equilibrium Bose-Einstein condensate (BEC) 
can deviate from the outcome of 
the time-dependent Gross-Pitaevskii dynamics, 
even in the limit of infinite number of particles and at
constant interaction parameter
when the system becomes 100\% condensed.
We demonstrate our finding on 
the dynamics of the center-of-mass position--momentum uncertainty product 
of a freely expanding as well as of a trapped BEC.
This time-dependent many-body phenomenon is explained 
by the existence of time-dependent correlations 
which manifest themselves in the system's 
reduced two-body density matrix used to evaluate the uncertainty product.
Our work demonstrates that one has to use a many-body propagation theory
to describe an out-of-equilibrium BEC, 
even in the infinite particle limit.
\end{abstract}


\maketitle 

\section{Introduction}\label{Intro}

The out-of-equilibrium dynamics of a quantum system 
is described by the time-dependent Schr\"odinger equation.
All physical information on the evolving quantum system 
can thus be obtained from its time-dependent wavefunction 
by applying various operators and calculating expectation values. 
The variance of an operator quantifies to what extent the system 
under investigation is in an eigenstate or a superposition of eigenstates of the operator. 
In this sense it dictates the quantum resolution by which the operator could be measured. 
The product of the variances of two operators defines an uncertainty product.
The uncertainty product quantifies to what extent two operators can be mutually measured.
As such, it is a fundamental concept in quantum mechanics.
A famous example is the position--momentum uncertainty product 
of a single quantum particle which is analyzed 
in quantum mechanics textbooks
for both the static and dynamic cases,
see, e.g., \cite{QM_book}.

Over the past two decades, since they were first experimentally realized \cite{ex1,ex2,ex3,NB1,NB2}, 
Bose-Einstein condensates (BECs) made of ultracold trapped bosonic atoms have become a popular ground to study interacting quantum systems, see the reviews \cite{rev1,rev2,rev3} and books \cite{book1,book2,book3}. 
There has been an intense theoretical interest in BECs,
and ample studies have been made 
to describe their static and particularly dynamic properties using Gross-Pitaevskii, mean-field theory. 
The time-dependent Gross-Pitaevskii equation governs a mean-field theory which
assumes that each and every boson is described by one and the same time-dependent one-particle function
throughout the evolution of the BEC in time.

The general paradigm is that Gross-Pitaevskii theory properly describes
the ground state as well as the out-of-equilibrium dynamics of BECs in the limit of large particle number.
To be explicit, 
Lieb, Seiringer, and Yngvason have rigorously proven for trapped BECs with two-body repulsive interaction,
in the limit of infinite particle number and at constant interaction parameter,
that the ground-state energy and density of the condensate converge 
to those obtained by minimizing the Gross-Pitaevskii energy functional \cite{Yngvason_PRA}.
Thereafter, Lieb and Seiringer proved in the same limit that
the ground state is 100\% condensed \cite{Lieb_PRL}. 
In the case of out-of-equilibrium dynamics,
Erd\H{o}s, Schlein, and Yau 
have rigorously proven that 
an expanding initially-trapped BEC, after the trap is released,
still exhibits 100\% condensation \cite{Erdos_PRL}.
Furthermore, the condensate density evolves according
to that predicted by the time-dependent Gross-Pitaevskii equation.

In a previous work \cite{Variance} we analyzed the {\it ground state} of a trapped BEC and demonstrated that, 
even in the infinite particle limit when the BEC is 100\% condensed, 
the variance of a many-particle operator can substantially differ from that predicted by the Gross-Pitaevskii theory. 
The existence of many-body effects beyond those predicted by the mean-field 
Gross-Pitaevskii theory stems from the necessity of performing the infinite particle limit only after the quantum mechanical observalbe is evaluated and not prior to its evaluation.
This is essential since otherwise any trace of many-body correlations is washed-out 
before the quantum mechanical observable can be evaluated.
This has been explained in length both analytically and numerically in Ref.~\cite{Variance}.

In the present work we generalize our previous result to the 
dynamics of an {\it out-of-equilibrium} BEC.
Dynamics is generally more intricate than statics,
and involves (many) excitations.
We show, analytically and numerically, 
that the evolution in time of the uncertainty product of two operators
can deviate from that of the Gross-Pitaevskii dynamics,
even in the infinite particle limit.
We explicitly demonstrate this deviation for the center-of-mass position--momentum uncertainty product 
of a freely expanding BEC as well as to the dynamics of a trapped BEC.
Our work advocates that one has to use a many-body propagation theory
to describe the out-of-equilibrium dynamics of BECs,
even in the limit of infinite number of particles
when the system becomes 100\% condensed.

The structure of the paper is as follows.
In Sec.~\ref{Ex1} 
we analyze the uncertainty product of a freely expanding, initially-trapped BEC. 
In Sec.~\ref{Ex2} we present a general theory
for the many-body uncertainty product in trapped BECs,
and show its importance for the case of the bosonic-Josephson-junction system.
Concluding remarks are put forward in Sec.~\ref{Conclusions}.

\section{The uncertainty product of a freely expanding Bose-Einstein condensate}\label{Ex1}

Even for the simplest case of an expansion from an harmonic trap,
one expects in the limit of infinite number of particles and at constant interaction parameter
the Gross-Pitaevskii and many-body dynamics to coincide.
Whereas following Ref.~\cite{Erdos_PRL} this indeed holds for the density of the BEC, 
as we shall see below this no longer is the case for the uncertainty product.
The expansion of a BEC from an harmonic trap has amply been discussed in the literature at the
mean-field level, see., e.g., Refs.~\cite{Castin_1996,Plaja_2002,Konotop_2003}.
Furthermore, the uncertainty product of an expanding BEC can be analyzed rather straightforwardly at the many-body level
due to the separability of the center-of-mass coordinate.
Yet, as far as we know,
contrasting at the infinite particle limit the exact time-dependent uncertainty product
and the Gross-Pitaevskii one
has not been made, which is the purpose of this section.
For the sake of completeness, 
we give a short derivation of the relevant many-body quantities below.

Consider the many-body Hamiltonian of $N$ interacting bosons in a three-dimensional trap $V_T(\r)$,
\beqn\label{HAM}
 \hat H_T(\r_1,\ldots,\r_N) &=& 
\sum_{j=1}^N \left[-\frac{1}{2} \frac{\partial^2}{\partial \r_j^2} 
+ \hat V_T(\r_j)\right] + \sum_{j<k} \lambda_0\hat W(\r_j-\r_k) = \nonumber \\
&=& \hat H(\r_1,\ldots,\r_N) + \sum_{j=1}^N \hat V_T(\r_j). \
\eeqn
Here, $\hbar=m=1$,
and $\hat W(\r_1-\r_2)$ is the boson-boson interaction with $\lambda_0$ its strength.
Here we only consider the case of repulsive interaction, i.e., $\lambda_0>0$.
The Cartesian components are $\r_j=(x_j,y_j,z_j)$,
$\frac{\partial }{\partial \r_j} = 
\left(\frac{\partial }{\partial x_j},\frac{\partial }{\partial y_j},\frac{\partial }{\partial z_j}\right)$.
The Hamiltonian $\hat H_T$ can in principle be time dependent,
although we will not exploit this option explicitly below.

The system is typically, but not necessarily, 
prepared in the ground state of the trap $V_T(\r)$,
\beq\label{GROUND}
\hat H_T(\r_1,\ldots,\r_N) \Phi(\r_1,\ldots,\r_N) = E \Phi(\r_1,\ldots,\r_N).
\eeq
The ground-state wavefunction $\Phi(\r_1,\ldots,\r_N)$ is normalized to unity.
We would like to examine the many-body dynamics when the system is released from the trap
\beqn\label{TDSE_PROP}
& & \hat H(\r_1,\ldots,\r_N) \Psi(\r_1,\ldots,\r_N;t) = i \frac{\partial\Psi(\r_1,\ldots,\r_N;t)}{\partial t} 
\quad \Longleftrightarrow \quad \nonumber \\
& & \quad \Longleftrightarrow \quad \Psi(\r_1,\ldots,\r_N;t) = e^{-i \hat H(\r_1,\ldots,\r_N) t} \Phi(\r_1,\ldots,\r_N). \
\eeqn
The initial condition is therefore $\Psi(\r_1,\ldots,\r_N;0)=\Phi(\r_1,\ldots,\r_N)$.

To discuss the dynamics it is useful to make a coordinate transformation,
from the laboratory frame $\r_1,\ldots,\r_N$ to 
the center-of-mass ({\it CM}) and relative-motion ({\it rel}) coordinates $\Q_1,\ldots,\Q_N$.
Explicitly, we define (see, e.g., Ref.~\cite{Cohen})
\beqn\label{COOR_NEW1}
& & \Q_k = \frac{1}{\sqrt{k(k+1)}} \sum_{j=1}^{k} (\r_{k+1}-\r_j), \qquad 1 \le k \le N-1, \qquad \nonumber \\
& & \Q_N = \frac{1}{\sqrt{N}} \sum_{j=1}^{N} \r_j. \
\eeqn
The corresponding conjugated momenta are $\frac{1}{i} \frac{\partial}{\partial \Q_k}$ where, in particular,
the momentum conjugated to $\Q_N$ equals
$\frac{1}{i} \frac{\partial}{\partial \Q_N}=\frac{1}{\sqrt{N}} \sum_{j=1}^{N} \frac{1}{i}\frac{\partial}{\partial \r_j}$.
Here, the respective Cartesian components are 
$\Q_k=(Q_{k,x},Q_{k,y},Q_{k,z})$ and $\frac{\partial }{\partial \Q_k} = 
\left(\frac{\partial }{\partial Q_{k,x}},\frac{\partial }{\partial Q_{k,y}},\frac{\partial }{\partial Q_{k,z}}\right)$.
With the help of the center-of-mass and relative-motion coordinates,
the Hamiltonian after the trapping potential is switched off becomes separable and reads
\beq\label{H_SEPAR}
\hat H(\Q_1,\ldots,\Q_{N-1},\Q_N) = -\frac{1}{2}\frac{\partial^2}{\partial \Q^2_N} + \hat H_{rel}(\Q_1,\ldots,\Q_{N-1}),
\eeq
where $\hat H_{rel}$ collects all other and only relative-motion terms.

We can now proceed to the time evolution of the expectation values of interest.
Let us look at the center-of-mass position and momentum operators in the $x$ direction
\beq\label{CM_X_P}
 \hat X_{CM} = \frac{1}{N} \sum_{j=1}^N \hat x_j, \qquad \hat P_{CM} = \sum_{j=1}^N \hat p_j,
\eeq
where $\hat p_j = \frac{1}{i}\frac{\partial}{\partial x_j}$.
They are proportional, respectively, to the $x$ component of $\Q_N$ and 
its conjugated momentum $\frac{1}{i} \frac{\partial}{\partial \Q_N}$,
i.e., $\hat X_{CM} = \frac{1}{\sqrt{N}} \hat Q_{N,x}$ and $\hat P_{CM} = 
\sqrt{N} \frac{1}{i} \frac{\partial }{\partial Q_{N,x}}$.
The center-of-mass position and momentum operators satisfy the commutation relation
$\bl\hat X_{CM},\hat P_{CM}\br=i$ for any number of bosons $N$.

We are interested in the evolution in time of the center-of-mass position--momentum uncertainty product,
thus we need the respective variances
\beqn\label{TD_VAR_X_P_2}
& & \Delta^2_{\hat X_{CM}}(t)=
\left[\langle\Psi(t)|\hat X_{CM}^2|\Psi(t)\rangle - \langle\Psi(t)|\hat X_{CM}|\Psi(t)\rangle^2\right] =
\Delta^2_{\hat X_{CM}}(0) + \nonumber \\
& & \quad + \left(\langle\Phi|\hat X_{CM} \hat P_{CM} + \hat P_{CM} \hat X_{CM}|\Phi\rangle 
- 2\langle\Phi|\hat X_{CM}|\Phi\rangle \langle\Phi|\hat P_{CM}|\Phi\rangle\right)\frac{1}{N} \cdot t +
\frac{\Delta^2_{\hat P_{CM}}(0)}{N^2} \cdot t^2, 
\nonumber \\
& & \Delta^2_{\hat P_{CM}}(t) = \left[\langle\Psi(t)|\hat P_{CM}^2|\Psi(t)\rangle - \langle\Psi(t)|\hat P_{CM}|\Psi(t)\rangle^2\right] = \Delta^2_{\hat P_{CM}}(0), \
\eeqn
where Eqs.~(\ref{TDSE_PROP},\ref{H_SEPAR},\ref{CM_X_P}) 
and the relations
$e^{+i \hat H t}\hat X_{CM}e^{-i \hat H t} = \hat X_{CM} + \frac{\hat P_{CM}}{N} t$ and 
$e^{+i \hat H t}\hat P_{CM}e^{-i \hat H t} = \hat P_{CM}$
have been used.
We see that the variances at time $t>0$ can be expressed by quantities at $t=0$ only.
This is a useful many-body relation.
Note that the term linear in time can vanish on the following accounts:
For the ground state in a general trap $\langle\Phi|\hat P_{CM}|\Phi\rangle=0$;
in a reflection-symmetric trap $\langle\Phi|\hat X_{CM}|\Phi\rangle=0$;
and in an harmonic trap, see for the example below,
$\langle\Phi|\hat X_{CM} \hat P_{CM} + \hat P_{CM} \hat X_{CM}|\Phi\rangle=0$.
Nonetheless, we keep this term to account for the most general case.
Finally, multiplying the center-of-mass variances in Eq.~(\ref{TD_VAR_X_P_2}) we get
\beqn\label{UNCER_CM_X_P}
& & \Delta^2_{\hat X_{CM}}(t) \Delta^2_{\hat P_{CM}}(t) = 
\Delta^2_{\hat X_{CM}}(0) \Delta^2_{\hat P_{CM}}(0) + \\
& & +\left(\langle\Phi|\hat X_{CM} \hat P_{CM} + \hat P_{CM} \hat X_{CM}|\Phi\rangle 
- 2\langle\Phi|\hat X_{CM}|\Phi\rangle \langle\Phi|\hat P_{CM}|\Phi\rangle\right) \frac{\Delta^2_{\hat P_{CM}}(0)}{N} \cdot t + \left[\frac{\Delta^2_{\hat P_{CM}}(0)}{N}\right]^2 \cdot t^2  \nonumber \
\eeqn
for their uncertainty product.

We are now in the comfortable position to examine a particular and solvable case of interest.
The fact that one can provide an analytical result is attractive and instrumental in what follows,
but by no means does it restrict the generality of the conclusions.
We consider a system prepared in the ground state of the harmonic potential $V_T(\r) = \frac{1}{2} \omega^2 \r^2$,
where $\omega$ is the frequency.
Since now $\Phi(\Q_1,\ldots,\Q_{N-1},\Q_N)=
\left(\frac{\omega}{\pi}\right)^{3/4} e^{-\frac{\omega}{2}\Q_N^2} \Phi_{rel}(\Q_1,\ldots,\Q_{N-1})$,
where $\Phi_{rel}$ depends on the relative-motion coordinates only,
the term linear in time of Eq.~(\ref{UNCER_CM_X_P}) vanishes
and the other terms at $t=0$ are readily evaluated and read
\beq\label{UNCER_CM_X_P_HO_1}
\Delta^2_{\hat X_{CM}}(0) = \frac{1}{2\omega N}, \qquad 
\Delta^2_{\hat P_{CM}}(0) = \frac{\omega N}{2}.
\eeq
The final result for the center-of-mass position--momentum uncertainty product of an
interacting system released out of the harmonic trap simplifies and is given by
\beq\label{UNCER_CM_X_P_HO_2}
\Delta^2_{\hat X_{CM}}(t) \Delta^2_{\hat P_{CM}}(t) = \frac{1}{4}\left(1 + \omega^2 t^2\right). 
\eeq
The result Eq.~(\ref{UNCER_CM_X_P_HO_2})
holds for any number of particles $N$ and general boson-boson interaction $\hat W(\r_1-\r_2)$.
In particular, it is the same behavior of a single free quantum particle
initially-prepared in a Gaussian wavefunction which is left to
propagate in time \cite{QM_book}.

We would like to contrast the analytical, many-body result from what one would obtain from the
Gross-Pitaevskii theory, which is expected to describe the exact out-of-equilibrium dynamics
in the limit of infinite number of particles \cite{Erdos_PRL}.
When the Gross-Pitaevskii equation,
\beq\label{TDGP}
 i \frac{\partial \phi_{GP}(\r;t)}{\partial t} = 
\left[ -\frac{1}{2}\frac{\partial^2}{\partial \r^2} + \lambda_0(N-1) \int d\r' \, 
 \phi^\ast_{GP}(\r';t) \hat W(\r-\r') \phi_{GP}(\r;t)\right] \phi_{GP}(\r;t),
\eeq
is employed to describe the system,
i.e., for $\Psi_{GP}(\r_1,\ldots,\r_N;t)=\prod_{j=1}^N \phi_{GP}(\r_j;t)$, 
the variances of the center-of-mass position and momentum operators read
\beqn\label{GP_VAR}
& & \Delta^2_{\hat X_{CM},GP}(t) =
\left[\langle\Psi_{GP}(t)|\hat X_{CM}^2|\Psi_{GP}(t)\rangle - \langle\Psi_{GP}(t)|\hat X_{CM}|\Psi_{GP}(t)\rangle^2\right] =
\nonumber \\
& & \qquad =
\frac{1}{N}\left[\langle\phi_{GP}(t)|\hat x^2|\phi_{GP}(t)\rangle 
- \langle\phi_{GP}(t)|\hat x|\phi_{GP}(t)\rangle^2\right], \nonumber \\
& & \Delta^2_{\hat P_{CM},GP}(t) =
\left[\langle\Psi_{GP}(t)|\hat P_{CM}^2|\Psi_{GP}(t)\rangle - \langle\Psi_{GP}(t)|\hat P_{CM}|\Psi_{GP}(t)\rangle^2\right] =
\nonumber \\
& & \qquad =
N\left[\langle\phi_{GP}(t)|\hat p^2|\phi_{GP}(t)\rangle 
- \langle\phi_{GP}(t)|\hat p|\phi_{GP}(t)\rangle^2\right]. \
\eeqn 
Consequently, the position--momentum uncertainty product at the mean-field level
of theory,
\beqn\label{GP_UNCER}
& & \left[\Delta^2_{\hat X_{CM}}(t) \Delta^2_{\hat P_{CM}}(t)\right]_{GP} =
\Delta^2_{\hat X_{CM},GP}(t) \Delta^2_{\hat P_{CM},GP}(t) = \\
& & = \left[\langle\phi_{GP}(t)|\hat x^2|\phi_{GP}(t)\rangle 
- \langle\phi_{GP}(t)|\hat x|\phi_{GP}(t)\rangle^2\right] \times 
\left[\langle\phi_{GP}(t)|\hat p^2|\phi_{GP}(t)\rangle 
- \langle\phi_{GP}(t)|\hat p|\phi_{GP}(t)\rangle^2\right], \nonumber \
\eeqn
depends explicitly on the Gross-Pitaevskii orbital.

Comparing Eqs.~(\ref{UNCER_CM_X_P_HO_2}) and (\ref{GP_UNCER})
implies that the time-dependent 
uncertainty product $\Delta^2_{\hat X_{CM}}(t) \Delta^2_{\hat P_{CM}}(t)$ computed at the many-body
level and at the Gross-Pitaevskii level cannot be the same.
This conclusion is based on the shape of the time-dependent Gross-Pitaevskii orbital $\phi_{GP}(\r;t)$. 
Since the Gross-Pitaevskii orbital is distorted due to the particle-particle
interaction from the shape of an expanding Gaussian 
(unless the interaction equals to zero), the position--momentum uncertainty product of any
time-dependent non-Gaussian one-particle function is different 
than the minimal uncertainty product $\frac{1}{4}\left(1 + \omega^2 t^2\right)$ of the exact result.
This is a generalization of the result for the ground state \cite{Variance}
to the out-of-equilibrium dynamics of a freely expanding, initially-trapped BEC.

\begin{figure}[!]
\includegraphics[width=0.8\columnwidth,angle=0]{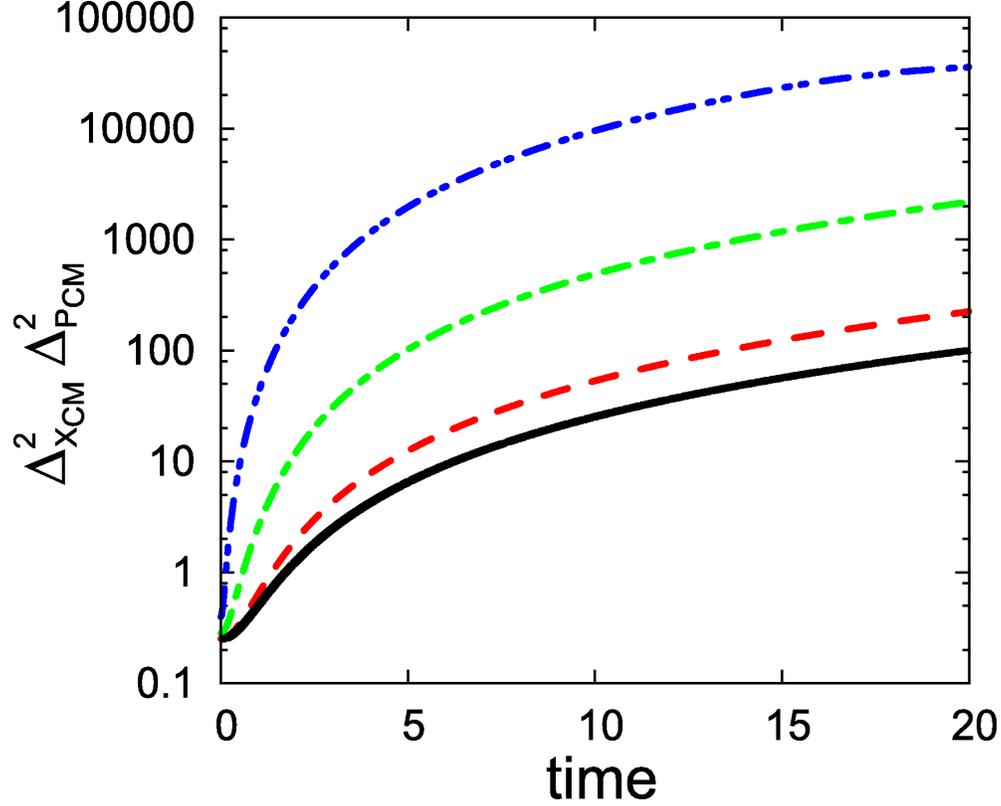}
\caption{(Color online) 
Time-dependent center-of-mass position--momentum uncertainty product
$\Delta^2_{\hat X_{CM}}(t) \Delta^2_{\hat P_{CM}}(t)$
of a BEC released from an harmonic trap.
An illustrative numerical example with one-dimensional bosons.
The one-body Hamiltonian is $-\frac{1}{2}\frac{\partial^2}{\partial x^2} + \frac{x^2}{2}$,
and the interboson interaction is contact, $\lambda_0\hat W(x_1-x_2)=\lambda_0\delta(x_1-x_2)$.
Shown and compared as a function of time $t$
are the Gross-Pitaevskii results for the interaction parameters $\Lambda = \lambda_0(N-1)=1$
(in red; dashed), $\Lambda=10$ (in green; dashed--dotted), and $\Lambda=100$ (in blue; dashed--double-dotted)
and the analytical, many-body result $\frac{1}{4}(1+t^2)$ valid $\forall \Lambda$ 
[Eq.~(\ref{UNCER_CM_X_P_HO_2})] 
(in black; full curve).
The position--momentum uncertainty product 
computed at the Gross-Pitaevskii level
differs from the analytical, many-body result.
The difference increases upon increasing $\Lambda$,
meaning that the pace of growth of the uncertainty at the mean-field level
depends on the interaction parameter $\Lambda$
(the y-axis is plotted in logarithmic scale).
The many-body uncertainty product grows as $t^2$,
and the mean-field uncertainty product is seen to grow in a similar manner in time
(see Ref.~\cite{Konotop_2003} for the mean-field analysis of the expansion).
The uncertainty product
constitutes a macroscopic probe of the time-dependent correlations of a BEC,
even when the system becomes 100\% condensed in the limit of infinite number of particles.
See the text for more details.
The quantities shown are dimensionless.
}
\label{f1_td}
\end{figure}

In order to explore the size of the deviation of the Gross-Pitaevskii dynamics from
the exact result we 
consider repulsive bosons in one spatial dimension
released from a harmonic potential.
The one-body Hamiltonian is $-\frac{1}{2}\frac{\partial^2}{\partial x^2} + \frac{x^2}{2}$,
and the boson-boson interaction is contact, $\lambda_0\hat W(x_1-x_2)=\lambda_0\delta(x_1-x_2)$.
Since there is no analytical solution to the time-dependent Gross-Pitaevskii equation, Eq.~(\ref{GP_UNCER}), 
we solve it numerically.
For the computation we have used $2000$ sine discrete-variable-representation grid points
in a box of size $[-100,100]$.
Figure~\ref{f1_td} collects the results.
In particular, the deviation
of the mean-field result (\ref{GP_UNCER}) 
from the analytical many-body relation (\ref{UNCER_CM_X_P_HO_2})
as a function of time is clearly visible.
The stronger the interaction parameter $\Lambda=\lambda_0(N-1)$ is,
the larger the deviation is.
This deviation can serve to quantify the amount of correlations in the system,
and how this amount evolves in time during the dynamics.  
The many-body uncertainty product increases like $t^2$ 
with the same pace of growth for any interaction strength, see Eq.~(\ref{UNCER_CM_X_P_HO_2}).
The mean-field uncertainty product is seen to increase similarly in time,
but the pace of growth depends on the interaction parameter $\Lambda$.
This is because the initially-trapped atomic cloud is broader than that of the non-interacting system 
and it expands faster with increasing $\Lambda$ \cite{Konotop_2003}.
All in all, 
the time-dependent, interaction-dressed shape of the atomic cloud
enters directly the computation of the uncertainty product 
$\Delta_{\hat X_{\mathrm CM}}^2(t)\Delta_{\hat P_{\mathrm CM}}^2(t)$
when it is performed at the mean-field level. 
We remind that this behavior occurs
even in the limit of infinite number of particles,
when the system becomes fully condensed.
With this example,
the analytical theoretical part of our work is concluded.
We move to the uncertainty product in the 
many-body dynamics of an out-of-equilibrium trapped BEC, 
for which a numerical solution and investigation 
are necessary.

\section{The uncertainty product in the dynamics of an out-of-equilibrium 
trapped Bose-Einstein condensate}\label{Ex2}

\subsection{Theory}\label{Theory}

The time-dependent Schr\"odinger equation of a trapped BEC,
\beq\label{TDSE}
\hat H_T(\r_1,\ldots,\r_N) \Psi(\r_1,\ldots,\r_N;t) = i \frac{\partial\Psi(\r_1,\ldots,\r_N;t)}{\partial t},
\eeq
where the wavefunction $\Psi(\r_1,\ldots,\r_N;t)$ is normalized to one,
generally has no analytical solution.
Thus, a numerical solution of the out-of-equilibrium dynamics is usually a must.
Furthermore and in the present context,
there is, to the best of our knowledge,
no mathematically-rigorous proof that the
out-of-equilibrium many-body Schr\"odinger dynamics of a trapped BEC 
is described -- in the limit of infinite number of particles and
at constant interaction parameter -- 
by the time-dependent Gross-Pitaevskii equation.
This is unlike the case of the ground state \cite{Yngvason_PRA,Lieb_PRL} or that of an expanding,
initially-trapped BEC \cite{Erdos_PRL}.
Thus, the additional merit of a many-body numerical treatment here
is to provide solid information on what can happen in this limit.

It is useful to proceed by employing the reduced one-body and two-body 
density matrices of $\Psi(\r_1,\ldots,\r_N;t)$ \cite{Cohen,Lowdin,Yukalov,Mazz,RDMs}.
The reduced one-body density matrix is given by
\beqn\label{1RDM}
\frac{\rho^{(1)}(\r_1,\r_1';t)}{N} &=&
\int d\r_2 \ldots d\r_N \, \Psi^\ast(\r_1',\r_2,\ldots,\r_N;t) \Psi(\r_1,\r_2,\ldots,\r_N;t) = \nonumber \\
 &=& \sum_j \frac{n_j(t)}{N} \, \alpha_j(\r_1;t) \alpha^\ast_j(\r'_1;t).
\eeqn
The quantities $\alpha_j(\r;t)$ are the so-called natural orbitals and $n_j(t)$ their respective occupations
which are time dependent and used to define the (possibly varying in time) 
degree of condensation 
in a system of interacting bosons \cite{Penrose_Onsager}.
The density of the system is the diagonal of the reduced one-body density matrix,
$\rho(\r;t) = \rho^{(1)}(\r,\r;t)$.

It is convenient to express in what follows quantities using the time-dependent natural orbitals $\alpha_j(\r;t)$.
The diagonal part of the reduced two-body density matrix is given by
\beqn\label{2RDM}
\frac{\rho^{(2)}(\r_1,\r_2,\r_1,\r_2;t)}{N(N-1)} &=& 
 \int d\r_3 \ldots d\r_N \, \Psi^\ast(\r_1,\r_2,\ldots,\r_N;t) \Psi(\r_1,\r_2,\ldots,\r_N;t) = \nonumber \\
 &=& \sum_{jpkq} \frac{\rho_{jpkq}(t)}{N(N-1)} \, 
\alpha^\ast_j(\r_1;t) \alpha^\ast_p(\r_2;t) \alpha_k(\r_1;t) \alpha_q(\r_2;t), 
\eeqn
where the matrix elements are 
$\rho_{jpkq}(t) = \langle\Psi(t)|\hat b_j^\dag \hat b_p^\dag \hat b_k \hat b_q|\Psi(t)\rangle$,
and the creation and annihilation operators are associated with the time-dependent natural orbitals.

We are interested in extracting the uncertainty product of operators 
for the time-dependent state of $N$ interacting bosons
in a trap described by the wavefunction\break\hfill $\Psi(\r_1,\ldots,\r_N;t)$.
We begin simply with the operator
\beq\label{lin_term}
 \hat A = \sum_{j=1}^N \hat a(\r_j)
\eeq
of the many-particle system,
where $\hat a(\r)$ is an hermitian operator.
A straightforward calculation gives 
the average per particle of $\hat A$ in the state $|\Psi(t)\rangle$,
$\frac{1}{N}\langle\Psi(t)|\hat A|\Psi(t)\rangle = \int d\r \frac{\rho(\r;t)}{N} a(\r)$,
which is 
seen to be directly related to the density of the system \cite{REM_XK}.

To compute the variance we also need the expectation value of the square of $\hat A$, 
\beq\label{sqr_term}
 \hat A^2 = \sum_{j=1}^N \hat a^2(\r_j) + \sum_{j<k} 2 \hat a(\r_j) \hat a(\r_k),
\eeq
which is comprised of one-body {\it and} two-body operators.
Expressed in terms of the above quantities, 
the time-dependent variance per particle of the operator $\hat A$ of the many-particle system reads
\beqn\label{dis}
\frac{1}{N}\Delta_{\hat A}^2(t) &=& \frac{1}{N} 
\left(\langle\Psi(t)|\hat A^2|\Psi(t)\rangle - \langle\Psi(t)|\hat A|\Psi(t)\rangle^2\right) = \nonumber \\
&=& \int d\r \frac{\rho(\r;t)}{N} a^2(\r) - \left[\int d\r \frac{\rho(\r;t)}{N} a(\r) \right]^2 + \nonumber \\
&+& \frac{\rho_{1111}(t)}{N} \left[\int d\r |\alpha_1(\r;t)|^2 a(\r) \right]^2 
- (N-1) \left[\int d\r \frac{\rho(\r;t)}{N} a(\r) \right]^2 + \nonumber \\
&+& \sum_{jpkq\ne 1111} \frac{\rho_{jpkq}(t)}{N} \left[\int d\r \alpha^\ast_j(\r;t) \alpha_k(\r;t) a(\r) \right]
\left[\int d\r \alpha^\ast_p(\r;t) \alpha_q(\r;t) a(\r)\right], \
\eeqn
where we reordered the terms as explained below. 
The first two describe the variance of $\hat a(\r)$ computed using the density per particle $\frac{\rho(\r;t)}{N}$.
The next two terms essentially cancel each other for condensed systems.
They exactly cancel each other within Gross-Pitaevskii theory,
since then $\rho_{1111}(t)=N(N-1)$ and $|\alpha_1(\r;t)|^2=|\phi_{GP}(\r;t)|^2 = \frac{\rho_{GP}(\r;t)}{N}$.
The last sum 
can contribute only when
a theoretical description beyond Gross-Pitaevskii is employed.
This allows the variance $\Delta_{\hat A}^2(t)$ 
to pick-up information from the other terms of the reduced two-body density matrix, $\rho_{jpkq}(t), jpkq\ne 1111$,
that normally do not contribute.
Furthermore, the explicit dependence of the last term in Eq.~(\ref{dis}) on integrals involving
the time-dependent natural orbitals $\{\alpha_k(\r;t)\}$ means that the contribution of 
many-body effects {\it depends} also on the instantaneous shape of the natural orbitals.
We will return to and examine this issue below.
Finally, in the absence of interboson interaction,
the time-dependent variance per particle $\frac{1}{N}\Delta_{\hat A}^2(t)$ boils down 
to that of a single particle.

It is useful for the sake of subsequent analysis to 
look at the difference in the variance of a system
when it is described at the many-body and Gross-Pitaevskii levels:
\beqn\label{dis_THERMO}
& & \frac{1}{N}\Delta_{\hat A}^2(t) - \frac{1}{N}\Delta_{\hat A, GP}^2(t) = \Delta_{\hat a, correlations}^2(t), \nonumber \\
& & \qquad \frac{1}{N}\Delta_{\hat A, GP}^2(t) = 
\int d\r \frac{\rho_{GP}(\r;t)}{N} a^2(\r) - \left[\int d\r \frac{\rho_{GP}(\r;t)}{N} a(\r)\right]^2. \
\eeqn
We henceforth call this difference the correlations term,
$\Delta_{\hat a, correlations}^2(t)$.
The correlations term vanishes identically for non-interacting bosons,
when all the bosons occupy one and the same time-dependent orbital.
For interacting bosons, 
if natural orbitals other than the first 
natural orbital (the so-called condensed mode) are at all occupied,
$\Delta_{\hat a, correlations}^2(t)$ does not vanish,
see also Eq.~(\ref{dis}).
To remind, the depletion of the BEC is given by the occupation of all but the first natural orbital divided by $N$.
For condensed systems the depletion is very small.
So, the question is whether a tiny occupation of the higher natural orbitals can make a difference?
We point out that in an out-of-equilibrium dynamics this occupation
may vary in time and still remain tiny. 
We already demonstrated in Sec.~\ref{Ex1} 
that the answer is positive for the freely expanding BEC.
We shall see below that the answer remains positive
for the dynamics of a trapped BEC as well.

Next, consider two many-particle operators, Eq.~(\ref{lin_term}), 
$\hat A=\sum_{j=1}^N \hat a(\r_j)$ and $\hat B=\sum_{j=1}^N \hat b(\r_j)$,
and their respective time-dependent variances per particle, 
$\frac{1}{N}\Delta_{\hat A}^2(t)$ and $\frac{1}{N}\Delta_{\hat B}^2(t)$.
The uncertainty product of these two operators satisfies the inequality \cite{REM_XK}
\beq\label{uncertainty_t_1}
\frac{1}{N}\Delta_{\hat A}^2(t) \frac{1}{N}\Delta_{\hat B}^2(t) 
\ge \frac{1}{4}\left|\int d\r \frac{\rho(\r;t)}{N} \bl\hat a(\r),\hat b(\r)\br\right|^2,
\eeq
where $\bl\hat A,\hat B\br = N \bl\hat a(\r),\hat b(\r)\br$
and $\bl\hat a(\r),\hat b(\r)\br$ is the usual commutator of the hermitian operators $\hat a(\r)$ and $\hat b(\r)$.
Using Eq.~(\ref{dis_THERMO}) we can express the difference
in the uncertainty product 
when it is described at the many-body and Gross-Pitaevskii levels as follows:
\beqn\label{uncertainty_t_3}
 & & \frac{1}{N}\Delta_{\hat A}^2(t) \frac{1}{N}\Delta_{\hat B}^2(t) 
     - \frac{1}{N}\Delta_{\hat A, GP}^2(t) \frac{1}{N}\Delta_{\hat B, GP}^2(t)
= \nonumber \\
& & \ \ = 
   \Delta_{\hat a, correlations}^2(t) \frac{1}{N}\Delta_{\hat B, GP}^2(t) +
   \Delta_{\hat b, correlations}^2(t) \frac{1}{N}\Delta_{\hat A, GP}^2(t) + \nonumber \\
& & \ \ + 
   \Delta_{\hat a, correlations}^2(t) \Delta_{\hat b, correlations}^2(t), \
\eeqn
We see that the difference in the uncertainty product Eq.~(\ref{uncertainty_t_3}) consists of contributions
from the correlations term in the variance of $\hat A$ times the mean-field variance of $\hat B$,
from the vice versa expression,
and from the product of the correlations terms in the variances of $\hat A$ and $\hat B$.
We will return to this analysis in the numerical study below.

\subsection{Case study: A bosonic Josephson junction}\label{Exam}

One of the most studied systems in the context of trapped BECs 
is the bosonic Josephson junction \cite{Milburn_1997,Smerzi_1997,Smerzi_1999,
Kasevich_2001,Vardi_2001,Markus_2005,Joerg_2005,Jeff_2007,Posa_2009,BJJ,Markus_2010,Universality,Shmuel_2015}.
We therefore would like to implement the above ideas and 
investigate numerically the time-dependent center-of-mass position--momentum uncertainty product
of a BEC in a double-well potential.
We need a suitable and proved many-body tool
to arrive at detailed conclusions.
Such a many-body tool is the multiconfigurational time-dependent
Hartree for bosons (MCTDHB) method,
which has been well documented \cite{MCTDHB1,MCTDHB2,book_MCTDH,book_nick},
benchmarked \cite{Benchmarks}, 
and extensively applied 
\cite{MCTDHB_OCT,MCTDHB_Shapiro,LC_NJP,Peter_2013,Tunneling_Con,
MCTDHB_3D_dyn,Breaking,Peter_2015a,Peter_2015b,Uwe,Sven_Tom,
Axel_ar,Alexej_ar,Kaspar_ar,Peter_ar}
in the literature.
Particularly, the numerically-exact quantum dynamics of the one-dimensional bosonic-Josephson-junction
system has been reported in Ref.~\cite{BJJ}.

\begin{figure}[!]
\includegraphics[width=0.8\columnwidth,angle=0]{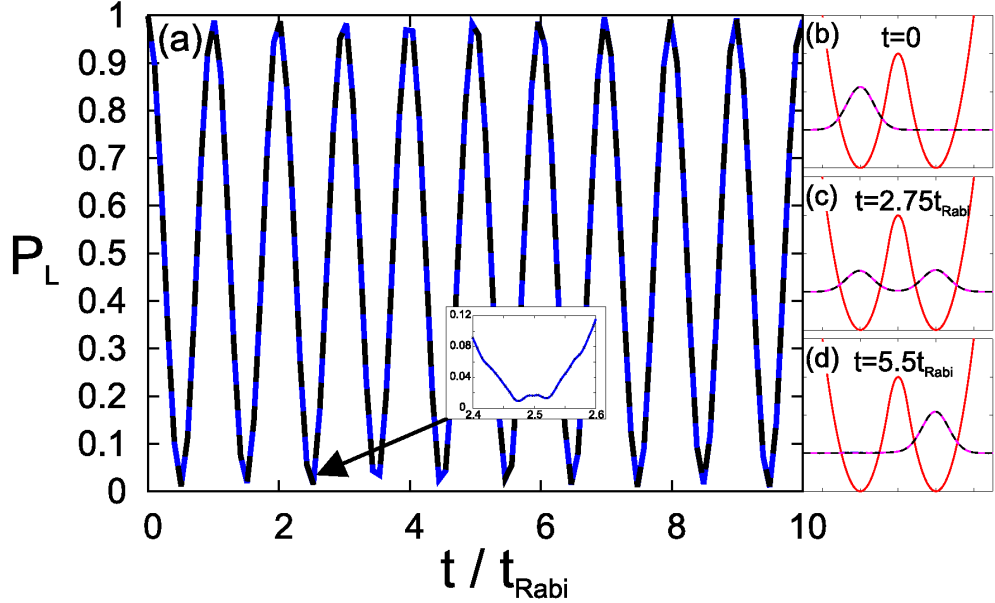}
\caption{(Color online) 
(a) Time-dependent survival probability in the left well, $P_L(t)$,
of a BEC in a one-dimensional bosonic Josephson junction.
The interboson interaction is contact, $\lambda_0\hat W(x_1-x_2)=\lambda_0\delta(x_1-x_2)$.
The interaction parameter is $\Lambda=\lambda_0(N-1)=0.01$.
Shown are the many-body results for $N=1\,000$, $10\,000$, and $100\,000$ bosons
(in color; full curves) using $M=2$ time-dependent orbitals,
and the Gross-Pitaevskii, mean-field result (in black; dashed curve).
All four curves, the three many-body results and the mean-field one,
lie on top of each other.
Full tunneling of the density back and forth 
between the left and right wells in seen.
The inset shows the smoothness of the survival probability at its extrema.
The panels to the right depict snapshots of the time-dependent density 
(with the double-well potential in the background) at three different times,
when the density is (b) either localized in the left well,
(c) delocalized over both wells,
(d) or localized in the right well.
The many-body (in color; full curves) and mean-field (in black; dashed curve) results
lie on top of each other.
That the many-body and mean-field results for the time-dependent density
and survival probability coincide 
indicates that the system is condensed (also see Fig.~\ref{f3_td}b).
See the text for more details.
The quantities shown are dimensionless.}
\label{f2_td}
\end{figure}

Consider a symmetric double-well potential $V_T(x)$ formed by fusing together
the two ``left'' and ``right'' harmonic potentials $\frac{1}{2}(x\pm 2)^2$ (see the appendix for details).
We begin by preparing at $t<0$ a BEC made of $N$ interacting bosons in the ground state of the left harmonic trap.
At $t\ge0$, the system is let to evolve in time in the double-well trap $V_T(x)$.
The bosons-boson interaction is contact, $\lambda_0\hat W(x_1-x_2)=\lambda_0\delta(x_1-x_2)$.
The interaction parameter is $\Lambda=\lambda_0(N-1)=0.01$.

\begin{figure}[!]
\vglue -1.6 truecm
\includegraphics[width=0.6\columnwidth,angle=0]{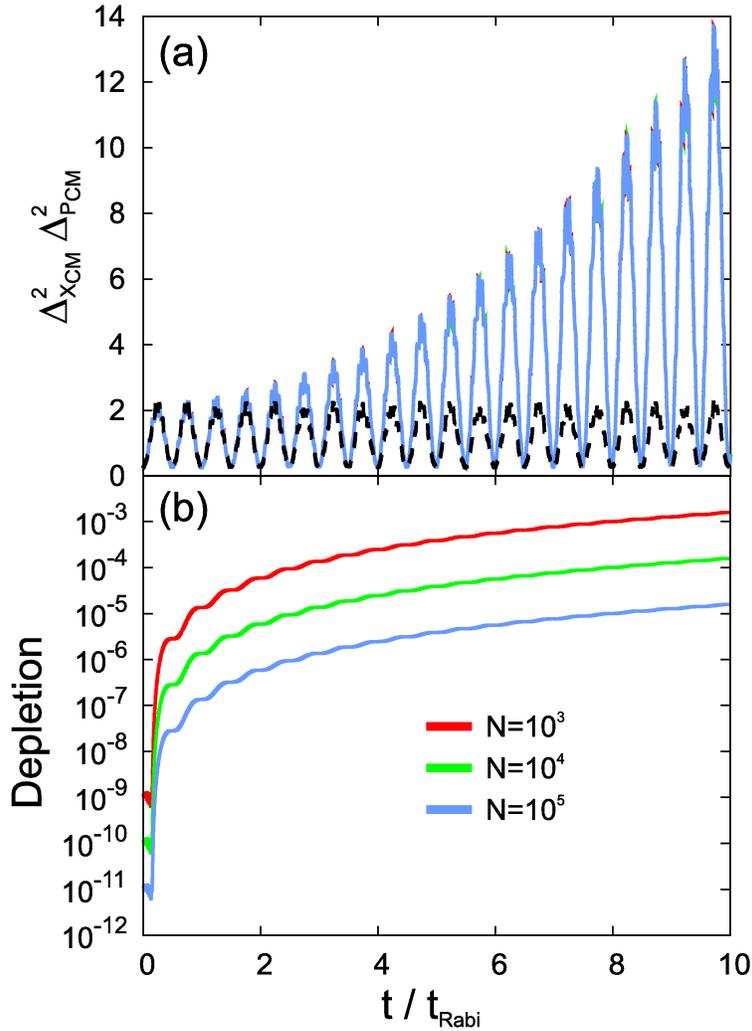}
\caption{(Color online) 
(a) Time-dependent center-of-mass position--momentum uncertainty product
$\Delta^2_{\hat X_{CM}}(t) \Delta^2_{\hat P_{CM}}(t)$
of a BEC in the one-dimensional bosonic Josephson junction
of Fig.~\ref{f2_td}.
Shown are many-body results using $M=2$ time-dependent orbitals.
The interaction parameter is $\Lambda=\lambda_0(N-1)=0.01$.
The system is weakly interacting.
The time-dependent uncertainty product computed at the many-body level 
is seen to grow in time in an oscillatory manner.
The period of oscillations is essentially $\frac{1}{2}t_{Rabi}$.
For comparison and later analysis, 
the Gross-Pitaevskii dynamics
only oscillates and thus seen to deviate from the many-body results as time progresses.
The results for $N=1\,000$ (in red), $10\,000$ (in green), and $100\,000$ (in light blue) bosons
lie on top of each other and suggest convergence of the many-body uncertainty product
with increasing number of bosons and at constant interaction parameter 
in the infinite particle limit.
(b) The time-dependent depletion decreases
 with $N$ and at constant $\Lambda$
(note the small values on the y axis which is in logarithmic scale). 
Side by side, the curves appear to be shifted vertically from each other,
suggesting that the total time-dependent number of particles
residing in the higher natural orbitals is constant.
See the text for more details.
The quantities shown are dimensionless.}
\label{f3_td}
\end{figure}

The time-dependent survival probability in the left well, 
$P_L(t) = \int_{-\infty}^{0} dx \frac{\rho(x;t)}{N}$,
of the BEC in the one-dimensional bosonic Josephson junction is registered in Fig.~\ref{f2_td}.
We compare the many-body time evolution of $N=1\,000$, $10\,000$, and $100\,000$ bosons
using $M=2$ time-dependent orbitals,
and the Gross-Pitaevskii, mean-field result ($M=1$).
It is found that all four curves, the three many-body calculations and the mean-field one,
lie atop each other.
We observe full tunneling of the density back and forth 
between the left and right wells. 
The slight deviation from the maximal (minimal) value of 100\% (0\%)
survival probability, and the small beating of the maxima of $P_L(t)$, 
are due to the finite depth of the double-well and the preparation of the initial condition in the left harmonic potential.
The apparent discontinuities are in fact smooth, see the inset in Fig.~\ref{f2_td}.
Important to our investigations is
that the many-body and mean-field results for the time-dependent density
and survival probability {\it coincide}.
This indicates that the system is fully condensed (also see Fig.~\ref{f3_td}b in this respect).
Furthermore, the interaction is so weak that known many-body effects in the bosonic Josephson junction,
such as the collapse of the density oscillations \cite{Milburn_1997,Universality},
have not yet developed.

Fig.~\ref{f3_td}a plots the time-dependent center-of-mass position--momentum uncertainty product
$\Delta^2_{\hat X_{CM}}(t) \Delta^2_{\hat P_{CM}}(t)$ in the bosonic Josephson junction \cite{CCC}.
The time-dependent uncertainty product computed at the many-body level 
is seen to grow in time in an oscillatory manner.
In contrast to the many-body result,
the Gross-Pitaevskii uncertainty product only oscillates and thus deviates 
from the many-body results as time progresses.
This represents a major difference between the mean-field
and many-body dynamics of the BEC.
Note that the results for $N=1\,000$, $10\,000$, and $100\,000$ bosons
lie on top of each other and suggest convergence of the many-body uncertainty product
with increasing number of bosons and at constant interaction parameter 
all the way to the infinite particle limit.

Along with the oscillatory growth of the uncertainty product,
the depletion of the (finite) BEC starts to increase, see Fig.~\ref{f3_td}b.
We observe that it is enough to have less than one particle
outside the condensed mode in order 
for the many-body uncertainty product to macroscopically differ from the mean-field one.  
Importantly, one can also see that the time-dependent depletion decreases
with increasing number of bosons $N$ and at constant interaction parameter $\Lambda$.
This suggests that the trapped time-dependent BEC remains 100\% condensed
in the limit of infinite number of particles.
Interestingly, the curves in Fig.~\ref{f3_td}b appear to be shifted vertically from each other,
suggesting that the total time-dependent number of particles
residing in the higher natural orbitals does not depend on $N$.
The latter time-dependent observation 
is analogous to the static situation for the ground state,
see Fig.~1c in Ref.~\cite{Variance}.

Let us analyze the dynamics of the uncertainty product more closely.
There are several features seen in Fig.~\ref{f3_td}a that we address:
The period of oscillations, the minimal and maximal values,
the fine structure atop the curves and, prominently,
the growth pace itself.
Since the BEC is fully condensed,
it is instrumental to first discuss the mean-field dynamics and then
the many-body one.
Because the repulsion is very weak, 
the period of oscillations of $P_L(t)$ is essentially 
that of the non-interacting system $t_{Rabi}$, see Fig.~\ref{f2_td}.
In each oscillations' cycle the BEC is localized once in the left and once in the right well.
Hence, the period of oscillations of the uncertainty product is half the period of 
the survival probability, i.e., $\frac{1}{2}t_{Rabi}$.

The minimal and maximal values of the center-of-mass uncertainty product can 
for the Gross-Pitaevskii dynamics be estimated as follows,
making use that the repulsion is very weak. 
When the BEC is localized in either the left or right well,
the minimal uncertainty product is essentially that of the 
initial-condition mean-field ground-state in the harmonic potential, 
i.e., $\Delta^2_{\hat X_{CM},GP}\Delta^2_{\hat P_{CM},GP}\approx \frac{1}{2N}\cdot\frac{N}{2}=\frac{1}{4}$.
On the other hand, when the BEC is delocalized over the two wells
and assuming their inter-well distance to be large
(the inter-well distance is equal to $4$),
the center-of-mass position variance is $\Delta^2_{\hat X_{CM}}\approx \frac{1}{N}\left(\frac{1}{2}+4\right)=\frac{9}{2N}$
and the momentum variance is still about $\Delta^2_{\hat P_{CM}}\approx \frac{N}{2}$.
Consequently, the maximal uncertainty product is 
$\Delta^2_{\hat X_{CM},GP}\Delta^2_{\hat P_{CM},GP}\approx \frac{9}{2N}\cdot \frac{N}{2}=\frac{9}{4}$.
These extreme values of the uncertainty product nicely
match the mean-field numerical results in Fig.~\ref{f3_td}.
The small deviations from these analytical values,
as well as the fine structure atop the maximal values of the uncertainty product, 
are due to the finite depth and size of the double well, which lead to a small mixture between the wells.
Consequently,
the initial condition prepared in the left harmonic well slightly
penetrates the right well,
and the sudden change at $t=0$ from a single to a double well
along with the
weak interaction mildly couple the initial wavepacket
to higher than the lowest two modes
of the non-interacting double well.

We are now in the position to discuss the many-body uncertainty.
At the very beginning of the density oscillations,
the mean-field and many-body uncertainty products lie atop each other, see Fig.~\ref{f3_td}a.
Then they start to deviate from each other in a unique manner.
The maxima of the many-body uncertainty product in the double well
starts to grow more and more in time (and the fine structure atop the curves is magnified), 
however the minima in the uncertainty product essentially coincide.
Also the period of oscillations are the same, see the respective survival probabilities in Fig.~\ref{f2_td}.
What is the reason for this unique difference?

Since the many-body and mean-field densities coincide and the system is condensed,
see Figs.~\ref{f2_td} and \ref{f3_td}b, respectively, 
the correlations term of the variance equals
essentially only the last sum in (\ref{dis}),
i.e., $\Delta_{\hat a, correlations}^2(t) \approx 
\sum_{jpkq\ne 1111} \frac{\rho_{jpkq}(t)}{N} \left[\int dx \alpha^\ast_j(x;t) \alpha_k(x;t) a(x) \right] \times $
$\left[\int dx \alpha^\ast_p(x;t) \alpha_q(x;t) a(x)\right]$.
Even the slightest occupation of the non-condensed modes, $\alpha_{k>1}(x;t)$,
see Fig.~\ref{f3_td}b,
is enough to generate a sizable value of the correlations term
and this macroscopically impacts the many-body variance, see Fig.~\ref{f3_td}a.

We are left now to analyze the individual contributions of
the position $\Delta_{\hat x, correlations}^2(t)$ and 
momentum $\Delta_{\hat p, correlations}^2(t)$ correlations terms
to the dynamics of the 
uncertainty product in the bosonic-Josephson-junction system.
A key point to note is that the density oscillates between the left and right wells back and forth.
Furthermore, the interaction is so weak that two time-dependent natural orbitals
are sufficient to converge the dynamics of the uncertainty product,
please see Fig.~\ref{f5_td} in the appendix.
When the density is localized in, say, the left well, 
both natural orbitals are localized in this well,
and the various contributions of the overlap integrals between the natural orbitals
times the respective elements of the reduced two-body density matrix
to the position and momentum correlations terms are small.
On the other hand,
when the density is spread over the two wells,
both natural orbitals,
the condensed $\alpha_1(x;t)$ and excited $\alpha_2(x;t)$ modes,
are delocalized over the junction and mimic
{\it gerade} and {\it ungerade} shapes.
Then, only the element 
$\frac{\rho_{1212}(t)}{N} \left[\int dx \alpha^\ast_1(x;t) \alpha_2(x;t) x \right]
\left[\int dx \alpha^\ast_1(x;t) \alpha_2(x;t) x\right]$
has a dominant contribution to the position correlations term.
Finally, examining the momentum correlations term,
the only possible dominant element,
$\frac{\rho_{1212}(t)}{N} \left[\int dx \alpha^\ast_1(x;t) \frac{1}{i}\frac{\partial}{\partial x}\alpha_2(x;t)\right]
\left[\int dx \alpha^\ast_1(x;t) \frac{1}{i}\frac{\partial}{\partial x} \alpha_2(x;t)\right]$, 
is, however, small as well.
This is because the delocalized natural orbitals look like linear combinations of Gaussian-like functions,
and taking their spatial derivatives leads to functions having nodes in each well, 
and thus small overlaps with the natural orbitals themselves.

Combining all the above,
when the density is spread over the two wells,
the elements of the reduced two-body density matrix times
the overlap integrals containing the natural orbitals give
a dominant many-body contribution to the uncertainty product.
Furthermore, analysis of the instantaneous shape of the time-dependent natural orbitals
traces this contribution to $\Delta_{\hat x, correlations}^2(t)$ rather than $\Delta_{\hat p, correlations}^2(t)$.
Interestingly, the many-body position variance surpasses the trap's size,
which sets in the maximal value of mean-field position variance.
This might be perceived as surprising,
because we are dealing with a condensed system confined in a trap.
But it is not, being a genuine many-body effect of the interacting particles.
Summing up,
the main contribution to the difference between the many-body and mean-field 
uncertainty product, see Eq.~(\ref{uncertainty_t_3}), in the bosonic Josephson junction dynamics
comes from the product of the time-dependent position correlations term
and the mildly changing mean-field variance of the momentum operator itself.
  
\begin{figure}[!]
\includegraphics[width=0.8\columnwidth,angle=0]{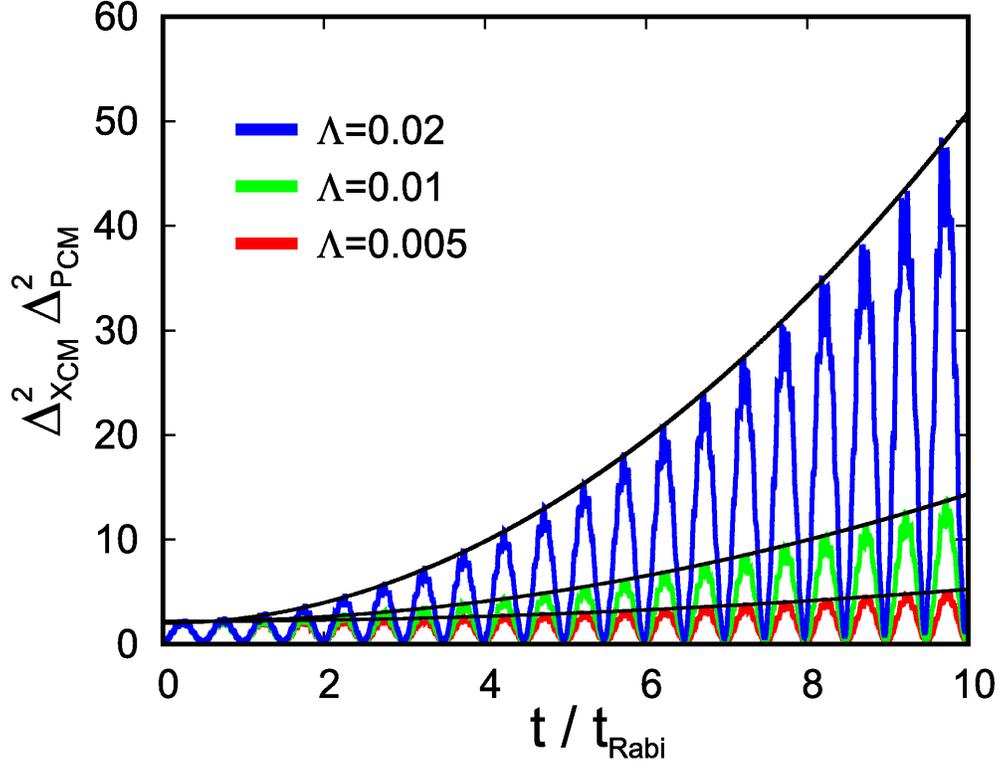}
\caption{(Color online) 
Time-dependent center-of-mass position--momentum uncertainty product
$\Delta^2_{\hat X_{CM}}(t) \Delta^2_{\hat P_{CM}}(t)$
of a BEC in the one-dimensional bosonic Josephson junction
of Fig.~\ref{f2_td}:
Analysis of its growth in time. 
Results for $N=1\,000$ bosons for the interaction parameters
$\Lambda=\lambda_0(N-1)=0.005$, $0.01$, and $0.02$ are depicted.
Shown are many-body results using $M=2$ time-dependent orbitals.
The black curves are least-squares fits
of the maxima of the uncertainty product
to a polynom of the form $a(\Lambda)+b(\Lambda) \cdot t^2$.
The numerical results suggest that the envelop of the time-dependent uncertainty product grows
as $\Lambda^2 t^2$, to leading orders in time and the interaction parameter.
See the text for more details.
The quantities shown are dimensionless.}
\label{f4_td}
\end{figure}

Finally, we set to examine the growth of the many-body uncertainty product in time.
In Fig.~\ref{f4_td} we have computed the many-body dynamics of $N=1000$ bosons
for the three interaction parameters $\Lambda=0.005$, $0.01$, and $0.02$.
The larger the interaction parameter is,
the faster the many-body uncertainty product grows (in an oscillatory manner).
We concentrate on the maxima values of the many-body uncertainty,
since at the minima the many-body and mean-field results
practically coincide.
A least-squares fit to a polynom of the form $a(\Lambda)+b(\Lambda) \cdot t^2$
was made to the maxima of each of the three curves in Fig.~\ref{f4_td}.
We can first see that the square growth in time nicely
matches the maxima values of the uncertainty.
We can numerically conclude that, to leading order in time, 
the (oscillating) many-body uncertainty product
of the BEC grows in the bosonic Josephson junction like $t^2$.

We can learn more from examining the 
interaction-dependent polynomial's coefficients $a(\Lambda)$ and $b(\Lambda)$.
The respective values are 
$a(0.005)=2.186$ and $b(0.005)=1.749 \cdot 10^{-6}$;
$a(0.01)=2.153$ and $b(0.01)=6.953 \cdot 10^{-6}$;
and
$a(0.02)=2.019$ and $b(0.02)=2.782 \cdot 10^{-5}$.
The baseline values are nearly the same,
and account for the many-body value of the uncertainty of the BEC, 
when it is momentarily delocalized between the two wells 
(extrapolated to the beginning of the propagation).
These values are
close to but slightly smaller
than the analytical value of $\frac{9}{4}$ discussed
above within the analysis of the mean-field uncertainty product. 
Furthermore, we find the following ratios for different interaction parameters:
$\frac{b(0.01)}{b(0.005)}=3.976 \approx 4$ and
$\frac{b(0.02)}{b(0.01)}=4.002 \approx 4$.
We can numerically conclude that, to leading order in the interaction parameter $\Lambda$, 
the (oscillating) many-body uncertainty product
of the BEC in the bosonic Josephson junction grows like $\Lambda^2$.
Consequently and all together, 
the many-body uncertainty product
in the double well grows to leading orders numerically like $\Lambda^2t^2$.
Finding an analytical explanation for this numerically-found growth law of the uncertainty product
would be interesting but goes beyond the scope of the present work.
We speculate that Ref.~\cite{Shmuel_2015}
could be generalized and instrumental here. 
Of course, finite BECs in the bosonic Josephson junction are interesting for their own right.
Then, as time goes by or when the interaction strength is increased,
known many-body effects, like the collapse of the density oscillations \cite{Milburn_1997,Universality}, 
eventually set in.
It would be interesting to explore the uncertainty product when
the system is no longer condensed.

\section{Concluding Remarks}\label{Conclusions}

We have discussed in the present work, analytically and numerically,
the time-dependent uncertainty product of an out-of-equilibrium BEC.
Although the systems considered are fully condensed, 
the uncertainty product at the many-body level differs from that predicted
at the Gross-Pitaevskii, mean-field level.
As the system evolves in time,
this difference can become substantial.
In this capacity,
our work demonstrates that one must use a many-body propagation theory
to describe the dynamics of an out-of-equilibrium BEC,
even in the limit of infinite number of particles
when the system becomes 100\% condensed.

To measure this time-dependent many-body effect,
the position and momentum of all particles in the clouds are in principle needed.
This is in contrast to the commonly reported particle density.
Experimental techniques which could achieve the necessary
resolution are constantly being improved \cite{exr1,exr2,exr3,exr4,exr5}.

\section*{Acknowledgements}

We thank Lorenz Cederbaum for discussions.
Computation time on the Cray XE6 system Hermit and
the Cray XC40 system Hornet 
at the HLRS are gratefully acknowledged.

\appendix

\section*{Appendix: Details of the time-dependent numerical computations}\label{APP_Num}

\begin{figure}[!]
\includegraphics[width=0.8\columnwidth,angle=0]{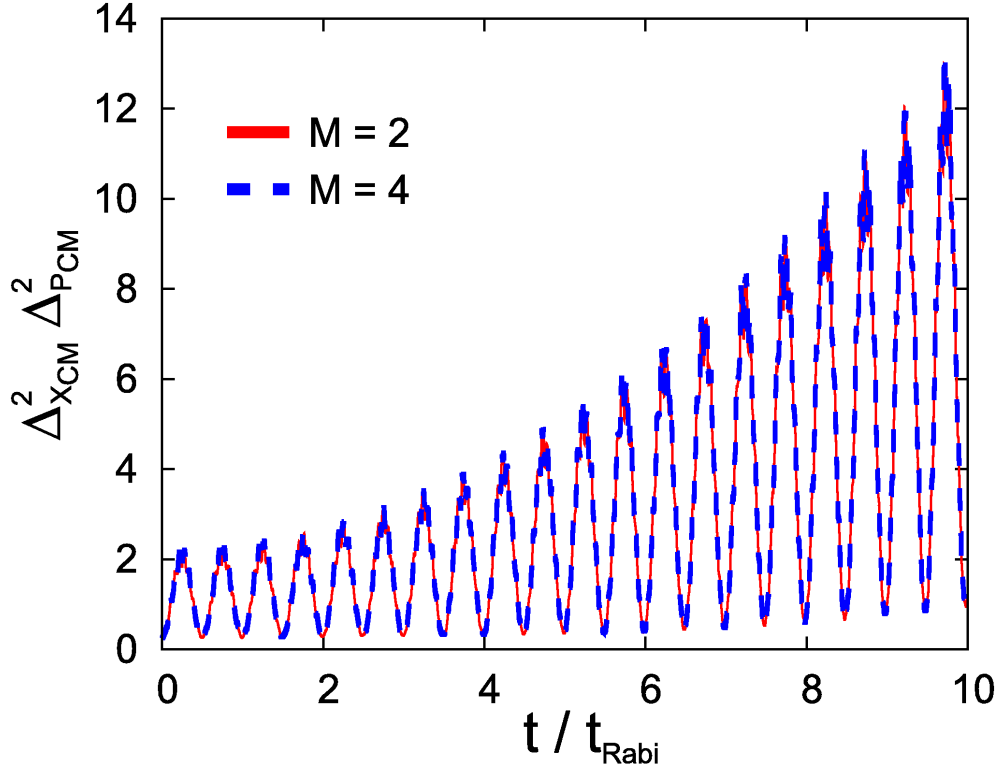}
\caption{(Color online) 
Time-dependent center-of-mass position--momentum uncertainty product
$\Delta^2_{\hat X_{CM}}(t) \Delta^2_{\hat P_{CM}}(t)$
of a BEC in the one-dimensional bosonic Josephson junction
of Fig.~\ref{f2_td}: Convergence of the out-of-equilibrium many-body dynamics.
Results for $N=100$ bosons for an increasing
number $M$ of the time-dependent orbitals are shown.
The curves lie on top of each other,
and the uncertainty product seen to converge with increasing $M$.
The quantities shown are dimensionless.}
\label{f5_td}
\end{figure}

The multiconfigurational time-dependent Hartree for bosons 
(MCTDHB) method \cite{MCTDHB1,MCTDHB2,book_MCTDH,book_nick} 
utilizes $M$ time-dependent orbitals, 
which are determined according to the Dirac-Frenkel, time-dependent variational principle,
to solve the time-dependent many-boson Schr\"odinger equation 
$\hat H_T(\r_1,\ldots,\r_N) \Psi(\r_1,\ldots,\r_N;t) = i \frac{\partial\Psi(\r_1,\ldots,\r_N;t)}{\partial t}$.
As the method is variational,
convergences in the limit $M \to \infty$ orbitals 
to the solution of the time-dependent many-boson Schr\"odinger equation is achieved \cite{Benchmarks}.
On the other end, i.e., for $M=1$,
the MCTDHB method boils down to the 
time-dependent Gross-Pitaevskii, mean-field theory.
The employment of $M$ time-dependent orbitals allows one to obtain accurate
numerical results with substantially less numerical resources
than a corresponding time-dependent computation employing $M$ fixed orbitals would need.
To obtain the initial conditions for the time propagations,
which here are the respective many-body ground states in the harmonic trap,
we propagate the MCTDHB equations-of-motion in imaginary time
\cite{Benchmarks,MCHB}. 
The MCTDHB method has recently been benchmarked 
and its accuracy demonstrated \cite{Benchmarks}. 
We use the numerical implementation in the software package \cite{package}.

In the present work the MCTDHB method is employed to compute the
time-dependent uncertainty product
in the out-of-equilibrium dynamics of a one-dimensional bosonic Josephson junction.
The many-body Hamiltonian is represented 
by $201$ harmonic-oscillator discrete-variable-representation grid points
in a box of size [-10,10].
The symmetric double-well potential 
is obtained by merging the two harmonic potentials
$\frac{1}{2}(x\pm2)^2$, whose minima are located, respectively,
to the left and right of the origin at $x=\mp2$, 
with a square-polynomial-barrier in the central region $|x|<\frac{1}{2}$,
giving rise to $V_T(x)=\big\{\frac{1}{2}(x+2)^2, x\le -\frac{1}{2}; \frac{3}{2}(1-x^2), 
|x|<\frac{1}{2}; \frac{1}{2}(x-2)^2, x\ge\frac{1}{2}\big\}$.
A convenient unit of time in the double well is the tunneling time,
$t_{\mathrm{Rabi}} = \frac{2\pi}{\Delta E} = 132.498$,
where $\Delta E$ is the energy difference between the ground state
and the first excited state of a single particle in the double-well potential.

As a concrete example and without loss of generality,
convergence with increasing $M$ 
of the time-dependent many-particle position--momentum uncertainty product
of an out-of-equilibrium BEC held 
in the one-dimensional bosonic Josephson junction of Fig.~\ref{f2_td} 
is demonstrated in Fig.~\ref{f5_td}.
Recall that the density and survival probability converge already and the Gross-Pitaevskii level
($M=1$), see Fig.~\ref{f2_td}. 
Fig.~\ref{f5_td} shows that
the dynamics of the many-particle position--momentum uncertainty product 
converges at the $M=2$ level of the MCTDHB theory.


\begin{thebibliography}{99}
\bibitem{REM_XK} This is in case of a local operator in coordinate space $\r$.
                          In the respective case in momentum space $\p$,  
                          $\frac{1}{N}\langle\Psi(t)|\hat A|\Psi(t)\rangle = \int d\p \frac{\rho(\p;t)}{N} a(\p)$.
                          For a `mixed' operator, like the angular momentum operator,
                          one generally requires its matrix elements with respect to the time-dependent natural orbitals.
\bibitem{CCC} Note that $\hat X_{CM}=\frac{\hat X}{N}$ and $\hat P_{CM}=\hat P$ implies that 
$\Delta_{\hat X_{CM}}^2(t) \Delta_{\hat P_{CM}}^2(t) =
\frac{1}{N}\Delta_{\hat X}^2(t) \frac{1}{N}\Delta_{\hat P}^2(t)$.
This means that the uncertainty product can be analyzed equivalently in terms of
the variances $\Delta_{\hat X_{CM}}^2(t)$ and $\Delta_{\hat P_{CM}}^2(t)$
or the variances per particle
$\frac{1}{N}\Delta_{\hat X}^2(t)$ and $\frac{1}{N}\Delta_{\hat P}^2(t)$.
\end{thebibliography}
\end{document}